\documentclass[12pt,epsf]{article}
\usepackage{amsmath,amssymb,graphicx,epsfig,latexsym}
\numberwithin{equation}{section}
\newcommand{\be}{\begin{equation}}
\newcommand{\ee}{\end{equation}}

\newcommand{\beqa}{\begin{eqnarray}}
\newcommand{\eeqa}{\end{eqnarray}}
\newcommand{\nn}{\nonumber}

\def\boxit#1{\vbox{\hrule\hbox{\vrule\kern8pt
\vbox{\hbox{\kern8pt}\hbox{\vbox{#1}}\hbox{\kern8pt}}
\kern8pt\vrule}\hrule}}
\def\mathboxit#1{\vbox{\hrule\hbox{\vrule\kern8pt\vbox{\kern8pt
\hbox{$\displaystyle #1$}\kern8pt}\kern8pt\vrule}\hrule}}

\def\IB{\relax\hbox{$\inbar\kern-.3em{\rm B}$}}
\def\IC{\relax\hbox{$\inbar\kern-.3em{\rm C}$}}
\def\ID{\relax\hbox{$\inbar\kern-.3em{\rm D}$}}
\def\IE{\relax\hbox{$\inbar\kern-.3em{\rm E}$}}
\def\IF{\relax\hbox{$\inbar\kern-.3em{\rm F}$}}
\def\IG{\relax\hbox{$\inbar\kern-.3em{\rm G}$}}
\def\IGa{\relax\hbox{${\rm I}\kern-.18em\Gamma$}}
\def\IH{\relax{\rm I\kern-.18em H}}
\def\IK{\relax{\rm I\kern-.18em K}}
\def\IL{\relax{\rm I\kern-.18em L}}
\def\IP{\relax{\rm I\kern-.18em P}}
\def\IR{\relax{\rm I\kern-.18em R}}
\def\IZ{\relax\ifmmode\mathchoice
{\hbox{\cmss Z\kern-.4em Z}}{\hbox{\cmss Z\kern-.4em Z}}
{\lower.9pt\hbox{\cmsss Z\kern-.4em Z}} {\lower1.2pt\hbox{\cmsss
Z\kern-.4em Z}}\else{\cmss Z\kern-.4em Z}\fi}

\def\II{\relax{\rm I\kern-.18em I}}

\def\CA {{\cal A}}

\def\CD {{\cal D}}

\def\CG {{\cal G}}

\def\CL {{\cal L}}

\def\CP {{\cal P}}

\def\CU {{\cal U}}

\begin{document}

\setlength{\baselineskip}{7mm}
\begin{titlepage}

\begin{flushright}

{\tt NRCPS-HE-27-2012} \\
{\tt CERN-PH-TH/2012-088}\\

\end{flushright}

\vspace{1cm}
%\begin{titlepage}
%\title{
\begin{center}

{\Large \it  New Gauge Anomalies \\
\vspace{0,3cm}
and
 \\
\vspace{0,3cm}
Topological Invariants in Various Dimensions \\
} %title ends

\vspace{1cm}
%\author{
{\sl Ignatios Antoniadis$^{~\star,}$\footnote{${}$ On leave of absence
from CPHT {\'E}cole Polytechnique, F-91128, Palaiseau Cedex,France.}
 and  George Savvidy$^{\star,+}$

\bigskip
\centerline{${}^\star $ \sl Department of Physics, CERN Theory Division CH-1211 Geneva 23, Switzerland}
\bigskip
\centerline{${}^+$ \sl Demokritos National Research Center, Ag. Paraskevi,  Athens, Greece}
\bigskip

}%author ends
%}
%\date{}%in order NOT to write the date
%\maketitle
\end{center}
\vspace{25pt}

\centerline{{\bf Abstract}}

\vspace{10pt}

\noindent
In the model of extended non-Abelian tensor gauge fields we have found
new metric-independent densities: the exact $(2n+3)$-forms  and their secondary
characteristics, the $(2n+2)$-forms as well as the exact
$6n$-forms and the corresponding secondary $(6n-1)$-forms.  These forms
are the analogs of the Pontryagin densities: the exact $2n$-forms and
Chern-Simons secondary characteristics, the $(2n-1)$-forms.
The $(2n+3)$- and $6n$-forms are gauge invariant densities,
while the $(2n+2)$- and $(6n-1)$-forms
transform non-trivially under gauge transformations, that we
compare with the corresponding transformations of the Chern-Simons secondary characteristics.
This construction allows to identify new potential gauge anomalies in various dimensions.
\\

\begin{center}
\end{center}

\vspace{150 pt}

%\end{abstract}
%\thispagestyle{empty}
%\end{titlepage}

\end{titlepage}

\newpage

\pagestyle{plain}
%\pagenumbering{roman}

\section{\it Introduction}

It is well known  that one can determine all chiral anomalies, Abelian
and non-Abelian \cite{Adler:1969gk,Bell:1969ts,Bardeen:1969md,Wess:1971yu,
Frampton:1983nr,Zumino:1983ew,Stora:1983ct,Faddeev:1984jp,Faddeev:1985iz,Mickelsson:1983xi,LBL-16443,
Manes:1985df,Treiman:1986ep,Faddeev:1987hg,DKFaddeev,AlvarezGaume:1985ex},
by a differential geometric method without having to evaluate the
Feynman diagrams.
The non-Abelian anomaly  in $2n$-dimensional space-time may be obtained  from
the Abelian anomaly in $2n+2$ dimensions by a series of reduction (transgression) steps
\cite{Zumino:1983ew,Stora:1983ct,Faddeev:1984jp,Faddeev:1985iz,LBL-16443,Manes:1985df,Treiman:1986ep,Faddeev:1987hg,AlvarezGaume:1985ex}.
The  reduction allows to construct topological densities which
are the non-Abelian anomalies and can be represented in a
compact integral form
\cite{Zumino:1983ew,Stora:1983ct,Faddeev:1984jp,Faddeev:1985iz,LBL-16443,Manes:1985df,
Treiman:1986ep,Faddeev:1987hg,AlvarezGaume:1985ex}.
The topological character of these densities has physical relevance
\cite{Belavin:1975fg,Callan:1976je,MIT-CTP-548,Witten:1983tw} and imposes
consistency restrictions on the quantum gauge field theories \cite{Bouchiat:1972iq,Gross:1972pv,Georgi:1972bb}.
It also provides  various topological mass-generation mechanisms in gauge theories
\cite{Deser:1982vy,Deser:1981wh,Schonfeld:1980kb}.
For instance, in the topologically massive gauge theory in three dimensions,
a Chern-Simons term included in the action makes gauge fields
massive~\cite{Deser:1982vy,Deser:1981wh,Schonfeld:1980kb}.
Furthermore, in a four-dimensional Abelian gauge field theory,
a topological entity called BF term
plays the a of a Chern-Simons term and generates a massive vector field
\cite{Cremmer:1973mg,Kalb:1974yc,Nambu:1975ba,Aurilia:1981xg,
Freedman:1980us,Allen:1990gb,Slavnov:1988sj,Dvali:2005ws}.
Generalization to the non-Abelian case
was recently suggested in \cite{arXiv:1001.2808}.

Our intension in this article is to extend these constructions to the
non-Abelian tensor gauge fields. Indeed, we  found two series of
invariant densities in various dimensions which are analogous
to the Pontryagin-Chern-Simons densities. First we shall review
the lower-dimensional case and then turn to the higher-dimensional
extensions.

In  the non-Abelian tensor gauge theory \cite{Savvidy:2005fi,Savvidy:2005zm,Savvidy:2005ki}
there exists a gauge invariant
metric-independent density $\Gamma(A) $ in five-dimensional
space-time\footnote{The definition of the higher-rank field-strength tensors
$G_{nm,q}$ is
given in eq.~(\ref{fieldstrengthparticular}) and we use Latin letters to numerate
five-dimensional coordinates $x_l$ $(l,m,n,..=0,1,\dots,4)$.} \cite{arXiv:1001.2808}:
\beqa\label{freeactionthreeprimesum}
\Gamma(A)  = \varepsilon^{lmnpq} ~Tr (G_{lm}G_{np, q})~
= \partial_{l} ~\Sigma^{l},
\eeqa
which is the derivative of the vector current $\Sigma_{l}$:
\be\label{topologicalcurrent1}
\Sigma^{l} = \varepsilon^{lmnpq} Tr (G_{mn }A_{pq}).
\ee
The current  $\Sigma^{l}$ is linear in the Yang-Mills (YM) field-strength
tensor $G_{mn }$ and in the rank-2 gauge field $A_{pq}$ which
has a symmetric and antisymmetric part, and only its antisymmetric part
involved in (\ref{topologicalcurrent1}). $G_{np, q}$ is
the field-strength tensor of the rank-2 gauge field (\ref{fieldstrengthparticular}).
The density $\Gamma(A)$ is diffeomorphism-invariant and does not involve the
space-time metric.
It is also invariant under the  group of gauge transformations
(\ref{globalcurv}):
$
 \Gamma(A^U)=\Gamma(A)  .
$
%The invariant $\Gamma(A)$ in five dimensions
It shares therefore many properties
of the Chern-Pontryagin density in
four-dimensional YM theory \cite{Belavin:1975fg,Treiman:1986ep}:
\be\label{chernpontragyn}
\CP(A)= {1\over 4}\varepsilon^{\mu\nu\lambda\rho }Tr G_{\mu\nu}G_{\lambda\rho }=
\partial_{\mu} C^{\mu},
\ee
which  is a derivative of the Chern-Simons topological vector current
\cite{Belavin:1975fg,Treiman:1986ep,Callan:1976je,MIT-CTP-548,Jackiw:1979ur,Jackiw:2004fg,Jackiw:2005wp}
\be\label{chernsimons}
C^{\mu}=\varepsilon^{\mu\nu\lambda\rho }Tr (A_{\nu}\partial_{\lambda}A_{\rho }
-i{2\over 3}g  A_{\nu}A_{\lambda}A_{\rho }).
\ee
Indeed, comparing the expressions (\ref{freeactionthreeprimesum}), (\ref{chernpontragyn})
and (\ref{topologicalcurrent1}), (\ref{chernsimons}) one can see that both entities $\CP(A)$
and $\Gamma(A)$ are
metric-independent,  insensitive to the local variation of the fields and are derivatives of the
corresponding vector currents $C^{\mu}$ and $\Sigma^{l}$. {\it The difference between them is that the former is
defined in four dimensions, while the latter in five.} This difference in one unit of the space-time
dimension originates from the fact that we have at our disposal high-rank tensor gauge fields
to build new invariants \cite{arXiv:1001.2808}.

While the invariant $\Gamma(A)$ and the vector current $\Sigma^{l}$ are defined on a five-dimensional
manifold, one can restrict the latter to a lower, four-dimensional manifold. The
restriction proceeds as follows.
Considering the fifth component of the vector current $ \Sigma^l$
\beqa\label{topologicalcharge}
 \varepsilon^{4nmpq} Tr (G_{nm }A_{pq})
\eeqa
 one can see that the remaining indices will not repeat
the external index and the sum is restricted to %the sum over
indices of four-dimensional space-time.
Therefore, we can reduce this functional to four dimensions, considering
%This is the case when the
gauge fields  independent
on the fifth coordinate $x_4$. This density  is  well
defined in four-dimensions %al space-time
and is gauge invariant under infinitesimal  gauge transformations up to
a total divergence term, as one can see below from (\ref{variation}).
Therefore we shall
consider its integral over four-dimensional space-time\footnote{We
are using Greek letters to numerate  the four-dimensional coordinates
$x_{\mu}$ ($\mu,\nu,\lambda,\dots=0,1,2,3)$.} \cite{arXiv:1001.2808}:
\beqa\label{topologicalSigmaCS}
\Sigma(A)
&=& {1 \over 32 \pi^2} \int_{M_4} d^4 x~\varepsilon^{\nu\lambda\rho\sigma} ~ Tr~ (G_{\nu\lambda } A_{\rho  \sigma}) .
\eeqa
This entity is an analog of the Chern-Simons integral\footnote{The $C^3$ component
of the topological current (\ref{chernsimons}) \cite{Jackiw:2004fg,Jackiw:2005wp,LBL-16443}.}
\be\label{chernsimonscharcterictic}
W(A) = {g^2\over 8 \pi^2} \int_{M_3}  d^3x ~\varepsilon^{ijk }~Tr~ (A_{i}\partial_{j}A_{k }
-i g {2\over 3} A_{i}A_{j}A_{k }),
\ee
but, {\it importantly, instead of being defined in three dimensions
it is  defined in four dimensions.} Thus, the non-Abelian tensor gauge fields
allow to build a natural generalization of the Chern-Simons characteristic in four-dimensional
space-time.

The functional $\Sigma(A)$ is invariant under infinitesimal  gauge transformations up to a total divergence term.
Indeed, its gauge variation under $\delta_\xi$, defined in (\ref{polygauge})-(\ref{fieldstrengthtensortransfor}), is
\beqa\label{variation}
\delta_{\xi} \Sigma(A)
&\propto& \varepsilon^{\nu\lambda\rho\sigma} \int_{M_4} Tr (-i g [G_{\nu\lambda }~\xi] A_{\rho  \sigma}
+ G_{\nu\lambda } (\nabla_{\rho} \xi_{\sigma}-i g  [A_{\rho  \sigma}~\xi] )) d^4 x =\nn\\
&=&\varepsilon^{\nu\lambda\rho\sigma} \int_{M_4} \partial_{\rho} ~Tr (
G_{\nu\lambda }  \xi_{\sigma} ) d^4 x=\varepsilon^{\nu\lambda\rho\sigma}
\int_{\partial M_4} Tr (
G_{\nu\lambda }  \xi_{\sigma} ) d\sigma_{\rho} =0 .
\eeqa
Here, the first and the third  terms cancel each other and the second one, after integration by
part and recalling the Bianchi identity  (\ref{bianchi}),  leaves only a
boundary term which vanishes  when
the gauge parameter $\xi_{\sigma}(x)$ tends to zero sufficiently fast at the boundary.
Hence, the functional is
invariant against {\it small} gauge transformations, but not under {\it large} ones for which
gauge transformations have a non-trivial behavior at the boundary.
Thus, we have to find out how $\Sigma(A)$
transforms  under large gauge transformations.
The expression we found has the form (\ref{largegaugetransformation}):
\beqa\label{largegaugetransformation0}
\Sigma(A^U) -\Sigma(A) = {i  \over 32 \pi^2 g}  \int_{ M_4}  d^4x
~\varepsilon^{\mu\nu\lambda\rho} \partial_{\lambda}  ~Tr~  (G_{\mu\nu}   U_{\rho} U^{-} ).
\eeqa
It reduces to (\ref{variation}) for the infinitesimal gauge transformations (\ref{small})
and allows to introduce a lower-dimensional density
\be\label{bababa}
\sigma^1_3(A,U) =
~\varepsilon^{ijk}  ~Tr~ (G_{ij}     U_{k} U^{-}  ) ~.
\ee
The expression (\ref{largegaugetransformation0})  is analogous to the corresponding one of
the Chern-Simons integral
\cite{Belavin:1975fg,Treiman:1986ep,Callan:1976je,MIT-CTP-548,
Jackiw:1979ur,Jackiw:2004fg,LBL-16443,Jackiw:2005wp}:
\beqa
 W(A^U) - W(A) &=& {1\over 8 \pi^2} \int_{M_3}  d^3x ~\varepsilon^{ijk }~
\partial_{i}~ Tr~ (\partial_{j}U U^{-} A_{k} ) \nn\\ &+&
{1\over 24 \pi^2} \int_{M_3}  d^3x ~\varepsilon^{ijk }~
Tr~ (U^{-}\partial_{i}U ~U^{-}\partial_{j}U ~U^{-}\partial_{k}U  )~,
\eeqa
and the density (\ref{bababa}) to the non-Abelian anomaly in two-dimensions
\cite{Zumino:1983ew,Stora:1983ct,Faddeev:1984jp,Faddeev:1985iz,LBL-16443}:
$$
\omega^{1}_{2}(A,U)=~\varepsilon^{jk} Tr(\partial_{j}U U^{-} A_{k} ).
$$

Indeed, the above consideration has deep relation with chiral anomalies appearing
in gauge theories interacting with Weyl fermions.
The Abelian $U_A(1)$ anomaly appears in
the divergence of the axial U(1) current $J^{A}_{\mu}= \bar{\psi} \gamma_{\mu}\gamma_5 \psi$,
and in four-dimensions %al space-time the    anomaly
it is given  by the divergence
\beqa\label{anomaly4dint}
\partial^{\mu} J^{A}_{\mu} = -{1\over 16 \pi^2} \varepsilon^{\mu\nu\lambda\rho } Tr (G_{\mu\nu}G_{\lambda\rho})
=-{1\over 4 \pi^2}\varepsilon^{\mu\nu\lambda\rho } \partial_{\mu} ~Tr (A_{\nu}\partial_{\lambda}A_{\rho }
-i{2\over 3}g  A_{\nu}A_{\lambda}A_{\rho }).
\eeqa
Similarly, the non-Abelian anomaly appears in the covariant divergence of the non-Abelian
left $J^{aL}_{\mu}= \bar{\psi}_L \gamma_{\mu}\gamma_5 \sigma^a\psi_L$  or right
$J^{aR}_{\mu}= \bar{\psi}_R \gamma_{\mu}\gamma_5 \sigma^a\psi_R$ handed currents, such as
\beqa\label{nonabeliananomalyint}
D^{\mu} J^{aL}_{\mu}
=-{1\over 24 \pi^2} \varepsilon^{\mu\nu\lambda\rho } \partial_{\mu} ~
Tr [\sigma^a (A_{\nu}\partial_{\lambda}A_{\rho }
-i{1\over 2}g  A_{\nu}A_{\lambda}A_{\rho })].
\eeqa
The Abelian anomaly is gauge invariant, while the non-Abelian anomaly is gauge covariant and is
given by the covariant divergence of a non-Abelian current.
These lower-dimensional densities have their higher-dimensional counterparts
\cite{Zumino:1983ew,Stora:1983ct,Faddeev:1984jp,Faddeev:1985iz,LBL-16443,Treiman:1986ep,Faddeev:1987hg,AlvarezGaume:1985ex}.
In $\CD = 2n$ dimensions,  the $U_A(1)$ anomaly is given by a $2n$-form,
the higher-dimensional analog of eq.~(\ref{anomaly4dint}):
\be
d*J^A ~\propto ~Tr(G^n)= d ~\omega_{2n-1},
\ee
where $\omega_{2n-1}$ is a generalization of the  Chern-Simons form to $2n-1$ dimensions
\cite{Zumino:1983ew,Treiman:1986ep}:
\be\label{integralformforAbelian}
\omega_{2n-1}(A)= n \int^1_0 d t ~ Tr(AG^{n-1}_t).
\ee
Here, we are using a shorthand notation  for the 2-form  YM field-strength tensor
$G=dA +A^2$ of the 1-form vector field $A = -ig A^{a}_{\mu} L_a dx^{\mu}$,
with $G_{t}= t G +(t^2-t)A^{2}$.

Our aim is to generalize  the above construction (\ref{freeactionthreeprimesum}), (\ref{topologicalSigmaCS}),
(\ref{largegaugetransformation0}) and (\ref{bababa}) by defining invariant densities in
higher dimensions $\CD=2n+3=5,7,9,11,\dots$~:
\be
\Gamma_{2n+3}(A)= Tr(G^n G_3) = d ~\sigma_{2n+2},
\ee
where we are using a shorthand notation  for the 3-form  field-strength tensor $G_3=dA_2 +[A,A_2]$
of the rank-2 gauge field
$A_2= -ig A^{a}_{\mu\nu} L_a dx^{\mu} \wedge dx^{\nu}$ and $G_{3t}=t G_3 +(t^2-t)[A,A_2]$.
The $(2n+2)$-form  $\sigma_{2n+2}$  is:
\be\label{sigma2n2}
\sigma_{2n+2}(A,A_2) =  \int^1_0 d t ~Tr(A  G^{n-1}_{t} G_{3t} +...+G^{n-1}_{t}  A G_{3t} + G^{n}_{t} ~ A_2).
\ee
Here, the 4-form $\sigma_4(A)$ coincides with the integrand of the functional (\ref{topologicalSigmaCS}).
In general, a $(2n+2)$-form $\sigma_{2n+2}(A)$ is defined in $\CD=2n+2=4,6,8,10,\dots$ dimensions.
The last equation is a generalization of the  Chern-Simons density in $2n-1$ dimensions
(\ref{integralformforAbelian}). The dimensionality of this density is $[mass]^{n(n+2)}$,
and it can be used as an addition
to the (2n+2)-dimensional Lagrangian density
\be
{1 \over F^{n^2 -2}} \int_{M_{2n+2}} \sigma_{2n+2}(A,A_2),
\ee
where F is a dimensional coupling constant, very similar to \cite{Witten:1983tw}.
The $n=1$ case $F \int_{M_{4}} \sigma_{4}$ was considered in
\cite{arXiv:1001.2808} as a gauge invariant mass generation mechanism.

We also found a second series of exact $6n$-forms constructed
only in terms of the 3-form gauge field-strength $G_3$:
\be
\Delta_{6n}= Tr(G_3)^{2n}= d \pi_{6n-1},
\ee
where for the $(6n-1)$-form one gets the following expression:
\be\label{pi6nform}
\pi_{6n-1}(A,A_2) = 2n \int^{1}_{0} ~d t ~ Tr (A_2 G^{2n-1}_{3t}).
\ee
These forms are defined in $\CD=6n-1=5,11,17,\dots$ dimensions.

As it was well understood in
\cite{Zumino:1983ew,Stora:1983ct,Faddeev:1984jp,Faddeev:1985iz,LBL-16443,Treiman:1986ep,Faddeev:1987hg,AlvarezGaume:1985ex},
the non-Abelian anomaly (\ref{nonabeliananomalyint}) is associated
with $\omega^{1}_{4}$, the gauge variation of the density
$\delta \omega_{5}= d \omega^{1}_{4}$  in (\ref{integralformforAbelian})
and with $\omega^1_{2n-2} $ in higher dimensions. Indeed,
a celebrated result for the non-Abelian anomaly
\cite{Zumino:1983ew,Stora:1983ct,Faddeev:1984jp,Faddeev:1985iz,LBL-16443,Treiman:1986ep,Faddeev:1987hg,AlvarezGaume:1985ex} can be obtained by
gauge variation of the $\omega_{2n-1}$:
\be\label{definition}
 \delta    \omega_{2n-1} = d \omega^1_{2n-2}~,
\ee
where the $(2n-2)$-form has the following integral representation \cite{Zumino:1983ew}:
\beqa\label{celebratedanomaly}
\omega^{1}_{2n-2}(\xi, A)=n(n-1) \int^{1}_{0} d t (1-t)~
Str \left(  \xi  d (A ~G^{n-2}_{t}  ) \right),
\eeqa
where $\xi = \xi^a L_a$ is a scalar gauge parameter and Str denotes a symmetrized trace.
In $\CD = 2n-2$ dimensions, the non-Abelian anomaly is given by this $(2n-2)$-form,  the
higher-dimensional analog of the equation (\ref{nonabeliananomalyint}):
\be
D*J^{L,R}_{\xi} ~\propto    ~\omega^{1}_{2n-2}(\xi,A).
\ee
Our next aim is to construct possible gauge anomalies $\sigma^{1}_{2n+1}$ and $\pi^{1}_{6n-2}$
which follow from the generalized densities
$\sigma_{2n+2}$ (\ref{sigma2n2}) and $\pi_{6n-1}$ (\ref{pi6nform}). These potential anomalies are defined through
the relation analogous to (\ref{definition}):
\be
\delta \sigma_{2n+2} = d \sigma^{1}_{2n+1},~~~~\delta \pi_{6n-1}=d \pi^{1}_{6n-2}.
\ee
The low-dimensional densities can be extracted  directly from (\ref{bababa})
and from (\ref{sigma2n2}). When we perform a vector-like gauge transformation  $\xi_1$,
where $\xi_1 = \xi^a_{\mu} L_a dx^{\mu}$ is a 1-form gauge parameter (\ref{polygauge}),
the corresponding densities are:
\beqa\label{a2a}
\sigma^{1}_{3}(\xi_1 , A) = Tr(\xi_1 G),~~~~~~
\sigma^1_5(\xi_1,A) = Tr(\xi_1 d(A d A + {1\over 2} A^3)),
\eeqa
and when the gauge transformation is performed by a scalar gauge parameter $\xi$, then
\be
\sigma^1_{5}(\xi,A,A_2) =Tr\left( \xi  ~d(~A dA_2  +A_2 dA + {1\over 2}  A^2 A_2 -{1\over 2}  A A_2 A
+ {1\over 2} A_2 A^2 ~) \right),
\ee
What is interesting
here is that $\sigma^1_{5}$  explicitly contains the second-rank gauge field $A_2$
when we perform the standard YM infinitesimal gauge transformation $\xi$.
Because it is defined in odd
dimensions it may have contribution to the parity-violating anomaly \cite{AlvarezGaume:1984nf} and
its descendant ($\delta \sigma^1_5 = d \sigma^2_4$)
\be\label{2cocycle}
\sigma^2_4(\xi,\eta,A) = Tr\left( (d \xi~ \eta + \eta~ d \xi -\xi ~d \eta - d\eta~ \xi )~ d A_2 \right)
\ee
may represent a potential  Schwinger term  in the corresponding gauge algebra   \cite{Faddeev:1984jp,Faddeev:1985iz}.

In the next section we shall present a short introduction into the theory of non-Abelian tensor
gauge fields and discuss their  small and large\footnote{The transformations that
are homotopic to the identity are called ``small", while those
that cannot be deformed to the identity are called
``large"\cite{Jackiw:2004fg,LBL-16443,Jackiw:2005wp}.} gauge transformations,
their field-strength tensors, the corresponding flat connections of the tensor gauge fields and their
Lagrangians \cite{Savvidy:2005fi,Savvidy:2005zm,Savvidy:2005ki}.
In section 3, we shall derive the expression (\ref{largegaugetransformation0})
for the  large gauge transformation  of the
functional $\Sigma(A)$ and compare it with the corresponding transformation of $W(A)$.
Equation (\ref{bababa}) for the descendant density $\sigma^{1}_{3}(A,U)$ will be
also presented. The corresponding integrands are the 4-form $\sigma_4$ (\ref{sigma2n2})
and 3-form $\sigma^{1}_{3}$ (\ref{a2a}).
In section 4, we  present a topological invariant in six dimensions
and its reduction to the 5-form $\pi_5$ and 4-form $\pi^{1}_{4}$. In
section 5, we shall derive general formulas for the $\sigma_{2n+2}$ given
in (\ref{sigma2n2}) and for the $\pi_{6n-1}$ of eq.~(\ref{pi6nform}).

In conclusions, we shall discuss and compare the Pontryagin-Chern-Simons densities
$\CP_{2n}$, $\omega_{2n-1}$ and $\omega^{1}_{2n-2}$ in
YM gauge theory with the corresponding two series of densities
$\Gamma_{2n+3}$, $\sigma_{2n+2}$, $\sigma^{1}_{2n+1}$
and $\Delta_{6n}$, $\pi_{6n-1}$ and $\pi^{1}_{6n-2}$  in the extended model containing
non-Abelian tensor gauge fields.  We shall also discuss different models suggested in the literature
describing the dynamics of an antisymmetric non-Abelian  tensor gauge field.
The existing no-go theorem \cite{Henneaux:1997mf}, which essentially limited
possible non-Abelian models, can be circumvent only if the model contains infinitely
many tensor gauge fields.

We have also included three appendices containing the basic formulas of tensor gauge fields that we use in the text and a short reminder on the winding number and non-Abelian anomaly.

\section{\it Small and Large Gauge Transformations  }

Let us shortly overview the model of massless tensor gauge fields
Lagrangian suggested in \cite{Savvidy:2005fi,Savvidy:2005zm,Savvidy:2005ki}.
The gauge fields are defined as rank-$(s+1)$ tensors
$$
A^{a}_{\mu\lambda_1 ... \lambda_{s}}(x),
$$
which are totally symmetric with respect to the
indices $  \lambda_1 \dots\lambda_{s}  $. The number of symmetric
indices $s$ runs from zero to infinity~\footnote{  {\it A priori} the tensor fields
have no symmetries with respect to the first index  $\mu$.}.
The index $a$ corresponds to the generators $L_a$
of an appropriate Lie algebra.
The extended non-Abelian gauge transformation $\delta_{\xi} $ (\ref{polygauge}),
(\ref{fieldstrengthtensortransfor}) of the tensor gauge fields
is defined in the Appendix  A and comprises a closed algebraic structure.
The generalized field-strength tensors
are defined as follows
\cite{Savvidy:2005fi,Savvidy:2005zm,Savvidy:2005ki}:
\beqa\label{fieldstrengthparticular}
G_{\mu\nu} &=&
\partial_{\mu} A_{\nu} - \partial_{\nu} A_{\mu} -
i g [A_{\mu}~A_{\nu}],\\
G_{\mu\nu,\lambda} &=&
\partial_{\mu} A_{\nu\lambda} - \partial_{\nu} A_{\mu\lambda} -
i g  (~[A_{\mu}~A_{\nu\lambda}] + [A_{\mu\lambda}~A_{\nu}] ~),\nn\\
G_{\mu\nu,\lambda\rho} &=&
\partial_{\mu} A_{\nu\lambda\rho} - \partial_{\nu} A_{\mu\lambda\rho} -
i g (~[A_{\mu}~A_{\nu\lambda\rho}] +
 [A_{\mu\lambda}~A_{\nu\rho}]+[A_{\mu\rho}~A_{\nu\lambda}]
 + [A_{\mu\lambda\rho}~A_{\nu}] ~),\nn\\[-7pt]
 &\cdot&\nn\\[-17pt]
 &\cdot&\nn\\[-17pt]
 &\cdot&\nn
% ......&.&............................................\nn
\eeqa
and transform homogeneously
with respect to the extended gauge transformations $\delta_{\xi} $.
The tensor gauge fields are in the matrix representation
$A^{ab}_{\mu\lambda_1 ... \lambda_{s}} =
(L_c)^{ab}  A^{c}_{\mu\lambda_1 ... \lambda_{s}} = i f^{acb}A^{c}_{\mu
\lambda_1 ... \lambda_{s}}$  with
$f^{abc}$ - the structure constants of the Lie algebra.

Using field-strength tensors one can construct  infinite series of forms
$
{{\cal L}}_{s}
$
invariant under the
transformations $\delta_{\xi} $. They are quadratic in field-strength
tensors. The
first terms are given by the formula
\cite{Savvidy:2005fi,Savvidy:2005zm,Savvidy:2005ki}:
\beqa\label{totalactiontwo}
{{\cal L}}=  {{\cal L}}_{YM}  +  {{\cal L}}_2 + ... =
&-&{1\over 4}G^{a}_{\mu\nu}G^{a}_{\mu\nu}\nn\\
&-&{1\over 4}G^{a}_{\mu\nu,\lambda}G^{a}_{\mu\nu,\lambda}
-{1\over 4}G^{a}_{\mu\nu}G^{a}_{\mu\nu,\lambda\lambda}\nn\\
&+&{1\over 4}G^{a}_{\mu\nu,\lambda}G^{a}_{\mu\lambda,\nu}
+{1\over 4}G^{a}_{\mu\nu,\nu}G^{a}_{\mu\lambda,\lambda}
+{1\over 2}G^{a}_{\mu\nu}G^{a}_{\mu\lambda,\nu\lambda}+...
\eeqa
The Lagrangian
contains quadratic in gauge fields kinetic terms, as well as cubic and
quartic terms  describing
non-linear interactions of gauge fields with dimensionless
coupling constant $g$.  The Lagrangian
$\CL$ is well defined in any dimension.

In studying topological properties of the extending Yang-Mills theory
it is  important to define finite (not infinitesimal) gauge
transformations of the tensor gauge fields. These can be
found by expansion of the ``large"   transformation of the gauge field $\CA_{\mu}(e)$
over the vector variable $e^{\mu}$ \cite{Savvidy:2005ki}. Thus
the large gauge transformation  of the tensor gauge fields takes the form
\beqa\label{global}
A^{U}_{\mu} &=& U^- A_{\mu} U + {i\over g}~U^{-}\partial_{\mu} U, \\
A^{U}_{\mu\lambda}&=& U^- A_{\mu\lambda} U + U^- A_{\mu} U_{\lambda}
-U^-U_{\lambda} U^- A_{\mu} U +
{i\over g}~(U^{-}\partial_{\mu} U_{\lambda} -U^-U_{\lambda} U^-\partial_{\mu} U),\nn\\[-7pt]
 &\cdot&\nn\\[-17pt]
 &\cdot&\nn\\[-17pt]
 &\cdot&\nn
%.......& &......................\nn
\eeqa
where $U_{\lambda} $ is the second term in the expansion of the unitary matrix
$\CU(\Xi(x,e))$  over the vector variable:
\beqa
 \CU(x,e) &=& U(x) + U_{\mu}(x)e^{\mu }+...,\nn\\
 \CU^{-}(x,e) &=& U^-(x) - U^-(x) U_{\mu}(x) U^-(x)e^{\mu }+...\nn
\eeqa
The field-strength tensors transform correspondingly:
\beqa\label{globalcurv}
G^U_{\mu\nu}&=& U^-G_{\mu\nu} U , \\
G^{U}_{\mu\nu,\lambda}&=& U^- G_{\mu\nu,\lambda} U + U^- G_{\mu\nu} U_{\lambda}
-U^-U_{\lambda} U^- G_{\mu\nu} U  ,\nn\\[-7pt]
 &\cdot&\nn\\[-17pt]
 &\cdot&\nn\\[-17pt]
 &\cdot&\nn
 %.......& &......................\nn
\eeqa
One can obtain the expressions for the large gauge transformations of the
higher-rank gauge fields $A^{a}_{\mu\lambda_1 ... \lambda_{s}}(x)$ by making further expansion of the unitary matrix
$\CU(\xi(x,e))$ over the  vector variable $e^{\mu}$.
In order to recover the infinitesimal gauge transformations (\ref{polygauge}) and
(\ref{fieldstrengthtensortransfor}) of the tensor
fields one should substitute the
infinitesimal form of the  matrices
\be\label{small}
U= 1 - i g L_a ~\xi^a(x),~~~U_{\mu}= -i g L_a ~\xi^a_{\mu}(x),...
\ee
into (\ref{global}) and (\ref{globalcurv}). As one can see, these matrix functions
provide a mapping
into the relevant gauge group G and into the corresponding algebra $\CG$.

Let us now find {\it the  flat connections,
that is, the  gauge field configurations which have non-trivial space-time behavior
and for which the corresponding field-strength
tensors (curvature) vanish}.
The YM field-strength $G_{\mu\nu }$ vanishes when the
vector potential is equal to a pure gauge connection:
\be\label{puregauge}
A^{flat}_{\mu}={i\over g}~U^{-}\partial_{\mu} U,
\ee
as it can be seen from the first equation in (\ref{global}). The higher-rank field-strength
tensor $G_{\mu\nu,\lambda }$ vanishes when the tensor field is
equal to the last term of the second equation in (\ref{global}):
\be\label{puregaugetensor}
A^{flat}_{\mu\lambda}={i\over g} ~(U^{-} \partial_{\mu} U_{\lambda} -
U^{-} U_{\lambda} U^{-} \partial_{\mu} U).
\ee
It is therefore a ``pure gauge connection" for the tensor gauge field.
{\it A  posteriori}  one can become  convinced  that $G_{\mu\nu,\lambda }$
indeed vanishes by calculating the field-strength tensor
(\ref{fieldstrengthparticular}) for the
field configurations (\ref{puregauge}) and (\ref{puregaugetensor}).

The physical states must be invariant under infinitesimal gauge transformation (\ref{small}),
or, equivalently,
under finite gauge transformations that are continuously connected to the
identity matrix $U=1$ and to the zero vector matrices  $U_{\mu}=0$.
But homotopically non-trivial gauge transformations that cannot be deformed to
the identity or to the zero vector matrices  may also be present.
The former %transformations that are homotopic to the identity and to the zero vector matrices  one
can  be called ``small"
in analogy with the standard YM theory \cite{Jackiw:1979ur,Jackiw:2004fg,Jackiw:2005wp,Treiman:1986ep}.
The gauge transformations
that cannot be deformed to the identity or to the zero vector matrices are called ``large".

Let us find the explicit form of the vector matrices $U_{\mu}$
for the $SU(2)$ gauge  group. Its group element $\CU(\Xi(x,e))$ can be parameterized
as follows:
\be\label{parametrization}
\CU= e^{-i   \Xi^a \sigma^a}= \cos  \vert \Xi \vert - i \hat{\Xi}^a \sigma^a \sin  \vert \Xi \vert,
\ee
where $\vert \Xi \vert =\sqrt{\Xi^a \Xi^a}$, ~$\hat{\Xi}^a  = \Xi^a / \vert \Xi \vert$ and
$\sigma^a$ are the Pauli matrices. Expanding the gauge parameter $\Xi^a(x,e)$ over the vector
variable $e^{\mu}$:
$$
\Xi^a(x,e) = \xi^a(x)+ \xi^a_{\mu}(x) e^{\mu}+...,
$$
where $\xi^a(x)$ and $ \xi^a_{\mu}(x)$ are space-time dependent gauge parameters,
one can get the required matrices $U_{\mu}$:
\be
U_{\mu}= {\partial \CU \over \partial \Xi^a } |_{e=0}~ \xi^a_{\mu}
=   -( \sin  \vert \xi \vert + i \hat{\xi}^a
\sigma^a \cos  \vert \xi \vert~)  ~\hat{\xi}^b \xi^b_{\mu}
-i \sigma^a (\delta^{ab}- \hat{\xi}^a \hat{\xi}^b )\xi^b_{\mu}~   {\sin   \vert \xi \vert ~
\over   \vert \xi \vert},
\ee
as well as the other matrix combinations appearing in the previous expressions:
\beqa
U^-U_{\mu}U^- &=&   ( \sin  \vert \xi \vert - i \hat{\xi}^a
\sigma^a \cos  \vert \xi \vert~)  ~\hat{\xi}^b \xi^b_{\mu}
-i \sigma^a (\delta^{ab}- \hat{\xi}^a \hat{\xi}^b )\xi^b_{\mu}~   {\sin   \vert \xi \vert ~,
\over   \vert \xi \vert} \\
U_{\mu}U^- &=&-i  \hat{\xi}^a \sigma^a~\hat{\xi}^b \xi^b_{\mu}
-i \sigma^a (\delta^{ab}- \hat{\xi}^a \hat{\xi}^b )\xi^b_{\mu}~   (\cos  \vert \xi \vert  + i \hat{\xi}^c
\sigma^c \sin  \vert \xi \vert~)~{\sin   \vert \xi \vert ~
\over   \vert \xi \vert},\nn\\
U^-U_{\mu} &=&-i  \hat{\xi}^a \sigma^a~\hat{\xi}^b \xi^b_{\mu} -i ( \cos  \vert \xi \vert + i \hat{\xi}^c
\sigma^c \sin  \vert \xi \vert~) \sigma^a (\delta^{ab}- \hat{\xi}^a \hat{\xi}^b )\xi^b_{\mu}~
{\sin   \vert \xi \vert \over   \vert \xi \vert},\nn
\eeqa
where $\vert \xi \vert =\sqrt{\xi^a \xi^a}$, ~$\hat{\xi}^a  = \xi^a / \vert \xi \vert$.
As usual, non-trivial boundary conditions can be imposed by requiring the following asymptotic behavior:
\be
 \vert \xi \vert ~~~{\over r \rightarrow \infty}> ~~~\pi N,
\ee
so that %in that case we have
\be
U ~~~{\over r \rightarrow \infty}> ~~~ \pm 1,~~~~~~~~~~
U_{\mu}~~~{\over r \rightarrow \infty}> ~~~\mp i  \hat{\xi}^a \sigma^a~
 \hat{\xi}^b \xi^b_{\mu} .
\ee

\section{\it Large Gauge Transformation of $\Sigma(A)$}

The infinitesimal gauge transformations of $\Sigma(A)$ defined in (\ref{topologicalSigmaCS})
can be expressed as total derivative (\ref{variation}), so that its
integral over four-dimensional space-time $M_4$ is given by contribution from the bounding surface and
vanishes  when the gauge parameter $\xi_{\sigma}(x)$ tends to zero sufficiently
fast at the boundary. We are interested to know how $\Sigma(A)$
transforms  when the gauge functions belong  to a non-trivial homotopy class
of unitary matrices.

From (\ref{puregauge}) and (\ref{puregaugetensor}) it follows that for our lower-rank
gauge fields {\it the variety  of non-trivial flat
connections are described by the  matrices $U$ and $U_{\lambda}$ }.
For a topological classification  it is required that U tends to a constant at infinity.
These gauge functions provide a mapping
into the relevant gauge group G, and for non-Abelian compact gauge groups such mappings
fall into disjoint homotopy classes labeled by an integer winding number  $\Pi^3(G) = Z$
\cite{Belavin:1975fg,Callan:1976je,MIT-CTP-548,Jackiw:1979ur,
Jackiw:2004fg,Jackiw:2005wp,Treiman:1986ep}.
The analytic expression for the winding number can be found by
considering how $W(A)$ of (\ref{chernsimonscharcterictic}) transforms under large
gauge transformations:
\beqa\label{largechernsimons}
W(A^U) = W(A) &+& {1\over 8 \pi^2} \int_{M_3}  d^3x ~\varepsilon^{ijk }~
\partial_{i} Tr~ (\partial_{j}U U^{-} A_{k} ) \nn\\ &+&
{1\over 24 \pi^2} \int_{M_3}  d^3x ~\varepsilon^{ijk }~
Tr~ (U^{-}\partial_{i}U ~U^{-}\partial_{j}U ~U^{-}\partial_{k}U  )~,
\eeqa
where the first term defines the non-Abelian anomaly in two-dimensions,
\be
\omega^1_2(A,U)= ~\varepsilon^{ij}~
Tr~ (\partial_{i}U U^{-} A_{j} )~,
\ee
which is the $n=2$ case of the general expression for anomaly (\ref{celebratedanomaly}),
and the last term is the  winding number  of the gauge
function $U$ (see Appendix B for a simple derivation)\footnote{ The term $
{1\over 8 \pi^2} \int_{M_3}  d^3x ~\varepsilon^{ijk }~
\partial_{i} Tr  ( \partial_{j}U ~U^{-}A_{k} ) $ does not contribute
for vector potentials $\vec{A}$ dropping  sufficiently fast at infinity,
$|\vec{A}| < 1/r$ \cite{LBL-16443,Jackiw:2005wp} and, in particular, for the pure gauge
connections (\ref{puregauge}).}.

There are many different ways to understand a topological charter and the meaning of the
functional $W(A)$. The above derivation is most suitable for our purposes.
Indeed,
in a similar way we would like to find out the transformation of the functional $\Sigma(A)$
when the gauge transformations are large.
Using the expression for $\Sigma(A)$
given in (\ref{topologicalSigmaCS}) and the transformation
laws (\ref{global}) and (\ref{globalcurv}) for the corresponding fields
one can derive the transformation of the functional in the following form:
\beqa
\Sigma(A^U) &=& {1 \over 32 \pi^2}\int_{ M_4}  d^4x
~\varepsilon^{\mu\nu\lambda\rho} ~Tr~ (U^- G_{\mu\nu} U)(U^- A_{\lambda \rho} U
+ U^- A_{\lambda } U_{\rho} -U^-U_{\rho} U^- A_{\lambda} U +\nn\\
&&~~~+{i\over g}~(U^{-}\partial_{\lambda} U_{\rho} -U^-U_{\rho} U^-\partial_{\lambda} U)).\nn
\eeqa
Using cyclic permutation of the matrices under the trace $Tr$ one can represent  it as
\beqa
\Sigma(A^U) &=& \Sigma(A) + {1 \over 32 \pi^2}\int_{ M_4}  d^4x
~\varepsilon^{\mu\nu\lambda\rho} ~Tr~ ( G_{\mu\nu}  [ A_{\lambda }, U_{\rho} U^- ]
+ {i\over g} G_{\mu\nu} \partial_{\lambda} (U_{\rho}U^-) ),\nn\\
\eeqa
and then combining the last two terms into a covariant derivative we get
\beqa
\Sigma(A^U) - \Sigma(A) &=& {i \over 32 \pi^2 g}  \int_{ M_4}  d^4x
~\varepsilon^{\mu\nu\lambda\rho} ~Tr~ \{G_{\mu\nu}  \nabla_{\lambda} (U_{\rho} U^{-} )\} \\
&=& {i \over 32 \pi^2 g}  \int_{ M_4}  d^4x
~\varepsilon^{\mu\nu\lambda\rho} ~Tr~ \{\nabla_{\lambda} (G_{\mu\nu}   U_{\rho} U^{-} )
-(\nabla_{\lambda} G_{\mu\nu} )  U_{\rho} U^{-} \}. \nn
\eeqa
By using Bianchi identity (\ref{bianchi}) the last expression can
be reduced to the following boundary integral:
\beqa
\Sigma(A^U) - \Sigma(A) &=& {i  \over 32 \pi^2 g}  \int_{ M_4}  d^4x
~\varepsilon^{\mu\nu\lambda\rho} ~Tr~ \{\nabla_{\lambda} (G_{\mu\nu}   U_{\rho} U^{-} )\}\nn\\
&=& {i  \over 32 \pi^2 g}  \int_{ M_4}  d^4x
~\varepsilon^{\mu\nu\lambda\rho} \partial_{\lambda}  ~Tr~  (G_{\mu\nu}   U_{\rho} U^{-} ) \nn\\
&=& {i \over 32 \pi^2 g}  \int_{ \partial M_4}
~\varepsilon^{\mu\nu\lambda\rho}  ~Tr~  (G_{\mu\nu}   U_{\rho} U^{-} )  ~d \sigma_{\lambda}.
\eeqa
Thus the final  expression for the large gauge transformation of $\Sigma(A)$
takes the form
\beqa\label{largegaugetransformation}
\Sigma(A^U) - \Sigma(A) &=& {i  \over 32 \pi^2 g}  \int_{ M_4}  d^4x
~\varepsilon^{\mu\nu\lambda\rho} \partial_{\lambda}  ~Tr~  (G_{\mu\nu}   U_{\rho} U^{-} ).
\eeqa
The expression (\ref{largegaugetransformation}) is analogous to the corresponding
formula for  the Chern-Simons integral (\ref{largechernsimons}) and
allows to introduce a lower-dimensional density
\be\label{secondf}
\sigma^1_3(A,U) =  ~\varepsilon^{ijk}  ~Tr~ (G_{ij}     U_{k} U^{-}  )~.
\ee
The gauge transformation of the $\sigma^1_3(A,U)$ does not generate a new density
and the reduction stops. As we shall see below, the higher-dimensional
density  $\sigma^1_5(A,U)$ transgresses further to $\sigma^2_4$ (\ref{2cocycle}).

\section{\it  Topological Density in Six Dimensions}
In the previous sections we considered the densities in five and four dimensions.
It is also possible to construct an invariant  in six dimensions. Let us
consider the following metric-independent  density in six dimensions:
\beqa\label{invarinatinsixdimensions}
\Delta(A)
= \varepsilon^{\mu\nu\lambda\rho\sigma\kappa} ~Tr ~G_{\mu\nu,\lambda}G_{\rho\sigma,\kappa },
\eeqa
which is  gauge invariant, because under infinitesimal gauge transformation $\delta_{\xi} $
(\ref{fieldstrengthtensortransfor}) its variation vanishes:
\beqa
\delta_{\xi} \Delta  &=& \varepsilon^{\mu\nu\lambda\rho\sigma\kappa}
Tr (\delta G_{\mu\nu,\lambda} G_{\rho\sigma, \kappa}+
G_{\mu\nu,\lambda} \delta G_{\rho\sigma, \kappa}) \nn\\
&=&-i g  \varepsilon^{\mu\nu\lambda\rho\sigma\kappa}
Tr ( [G_{\mu\nu,\lambda} ~\xi] +[G_{\mu\nu} ~\xi_{\lambda}]) G_{\rho\sigma,\kappa}
 +G_{\mu\nu,\lambda} (~[~G_{\rho\sigma,\kappa}~ \xi ]
+  [G_{\varrho\sigma} ~\xi_{\kappa}]~))=0.\nn
\eeqa
Note that in the presence of a non-vanishing gravitational
background $R_{\mu\nu \lambda\rho}$  the transformation of the field-strength
tensors changes and instead of (\ref{fieldstrengthtensortransfor}) has the following form:
\beqa\label{fieldstrengthtensortransfor1}
\delta G_{\mu\nu}&=&  -i g  [G_{\mu\nu} ~\xi] , \\
\delta G_{\mu\nu,\lambda} &=& -i g  (~[~G_{\mu\nu,\lambda}~ \xi ]
+  [G_{\mu\nu} ~\xi_{\lambda}]~) + 2 R_{\mu\nu \lambda\rho} \xi^{\rho},\nonumber\\[-7pt]
 &\cdot&\nn\\[-17pt]
 &\cdot&\nn\\[-17pt]
 &\cdot&\nn
% ......&.&............................................\nn
\eeqa
while $\Delta(A) $ remains gauge invariant in that case
as well. The variation gets an additional term
$ \varepsilon^{\mu\nu\lambda\rho\sigma\kappa}
Tr ( G_{\mu\nu,\lambda} R_{\rho\sigma  \kappa \alpha}\xi^{\alpha})$,
which is equal to zero if one uses the permutation property of the Riemann curvature tensor:
$$
R_{\rho\sigma  \kappa \alpha} + R_{\sigma  \kappa\rho \alpha} +R_{\kappa \rho\sigma  \alpha}=0.
$$
The density $\Delta(A) $ does not involve the space-time metric and is diffeomorphism invariant.
Moreover, it is  invariant not only under infinitesimal gauge
transformations $\delta_{\xi} $, but also under large  transformations (\ref{globalcurv}):
\be
\Delta(A^U)=\Delta(A).
\ee
The density $\Delta(A) $ is a total derivative of a
vector current $\Pi_{\mu}$:
\beqa\label{exactformhigh}
\Delta(A)
=\varepsilon^{\mu\nu\lambda\rho\sigma\kappa} ~
Tr ~G_{\mu\nu,\lambda}G_{\rho\sigma, \kappa}=
2 ~\partial_{\mu} \Pi^{\mu},
\eeqa
where
\beqa\label{topologicalcurrenthigh}
\Pi^{\mu}(A)
&=&   \varepsilon^{\mu\nu\lambda\rho\sigma\kappa} ~Tr ~ G_{\nu\lambda,\rho }  A_{\sigma \kappa }.
\eeqa
While $\Delta(A) $ and the vector current $\Pi_{\mu}(A)$ are defined on a six-dimensional
manifold, we may restrict the latter to a lower, five-dimensional manifold.
Considering, indeed, the sixth component of the vector current $\Pi_{\mu}$
\beqa\label{topologicalchargehigh}
 \varepsilon^{5\nu\lambda\rho\sigma\kappa} ~Tr G_{\nu\lambda,\rho   }A_{\sigma \kappa}.
\eeqa
one sees that the remaining indices do not repeat
the external index and the sum is restricted to five-dimensional indices.
One can thus reduce this functional to five dimensions, considering
gauge fields independent
of the sixth coordinate $x_5$. The density $\Pi^5$ is then well
defined in five-dimensional space-time and, as we shall see,
it is also gauge invariant up to a total divergence term. We can therefore
consider its integral over five-dimensional space-time:
\beqa\label{topologicalSigmaCShigh}
\Pi(A)
&=&  \varepsilon^{\nu\lambda\rho\sigma\kappa} \int_{M_5} d^5 x~ Tr~ G_{\nu\lambda, \rho  } A_{\sigma\kappa} .
\eeqa
This functional is gauge invariant up to a total divergence term.
Its infinitesimal gauge variation under $\delta_\xi$
(\ref{polygauge})-(\ref{fieldstrengthtensortransfor}) is given by
\beqa\label{variationhigh}
\delta_{\xi} \int_{M_5} d^5 x~ \Pi
=\varepsilon^{\nu\lambda\rho\sigma\kappa} \int_{M_5} \partial_{\sigma} ~Tr (
G_{\nu\lambda,\rho }  ~\xi_{\kappa} ) d^5 x=\varepsilon^{\nu\lambda\rho\sigma\kappa}
\int_{\partial M_5} Tr (
G_{\nu\lambda, \rho}  \xi_{\kappa} ) d\sigma_{\sigma} =0 ,\nn
\eeqa
where the boundary term vanishes  when
the gauge parameter $\xi_{\kappa}$ tends to zero at infinity. But  it changes
under large gauge transformations. Let us calculate its variation.

Using the  transformation
properties (\ref{global}) and (\ref{globalcurv}) of the corresponding fields,
one can derive the transformation of the functional in the following form:
\beqa
\Pi(A^U)  &=&  \int_{ M_5}  d^5x
~\varepsilon^{\mu\nu\lambda\rho\sigma} ~Tr~ (U^- G_{\mu\nu,\lambda} U
+  U^- G_{\mu\nu} U_{\lambda} - U^-U_{\lambda} U^-G_{\mu\nu } U)\nn\\
&&(U^- A_{ \rho \sigma} U
+ U^- A_{\rho } U_{\sigma} -U^-U_{\sigma} U^- A_{\rho} U +
 {i\over g}~(U^{-}\partial_{\rho} U_{\sigma} -U^-U_{\sigma} U^-\partial_{\rho} U)).\nn
\eeqa
Using cyclic permutation of the matrices under the trace, one can represent  it as
\beqa
\Pi(A^U) &=& \Pi(A) +  \int_{ M_5}  d^5x
~\varepsilon^{\mu\nu\lambda\rho\sigma} ~Tr~\{ G_{\mu\nu,\lambda}  [ A_{\rho }, U_{\sigma} U^- ]
+ {i\over g} G_{\mu\nu,\lambda} \partial_{\rho} (U_{\sigma}U^- )  + \nn\\
&+&G_{\mu\nu } U_{\lambda}U^- [ A_{\rho }, U_{\sigma} U^- ]
+ {i\over g} G_{\mu\nu}  U_{\lambda}U^- (\partial_{\rho} U_{\sigma}U^-)-\nn\\
&-&G_{\mu\nu } [ A_{\rho }, U_{\sigma} U^- ] U_{\lambda}U^-
- {i\over g} G_{\mu\nu} (\partial_{\rho} U_{\sigma}U^-)U_{\lambda}U^-\nn\\
&+&G_{\mu\nu } [U_{\lambda}U^-  A_{\rho \sigma }]
- {i\over g} G_{\mu\nu} (\partial_{\rho} U_{\sigma}U^-)U_{\lambda}U^-
+[A_{\rho \sigma } G_{\mu\nu }  ] U_{\lambda}U^-  \}\nn
\eeqa
and then, combining the  terms into the covariant derivative, we get
\beqa
&& \Pi(A^U) - \Pi(A) = \nn\\
&=&  {i \over g} \int_{ M_5}  d^5x
~\varepsilon^{\mu\nu\lambda\rho \sigma} ~
Tr~ \{G_{\mu\nu,\lambda}  \nabla_{\rho} (U_{\sigma} U^{-} )
-ig [A_{\rho \sigma } G_{\mu\nu }  ]U_{\lambda} U^{-} +
 G_{\mu\nu }  \nabla_{\rho} (U_{\lambda} U^{-}   U_{\sigma} U^{-} ) \} \nn\\
&=&  \int_{ M_5}  d^5x
~\varepsilon^{\mu\nu\lambda\rho\sigma} ~Tr~ \{\nabla_{\rho} (G_{\mu\nu,\lambda}   U_{\sigma} U^{-} )
-(\nabla_{\rho} G_{\mu\nu,\lambda} )  U_{\sigma} U^{-}
+ig [A_{\rho \lambda } G_{\mu\nu }  ]U_{\sigma} U^{-} +\nn\\
& & ~~~~~~~~~~~~~~~~~~~~~~~~~+\nabla_{\rho} (G_{\mu\nu } U_{\lambda} U^{-}   U_{\sigma} U^{-}  ) -
(\nabla_{\rho} G_{\mu\nu } )U_{\lambda} U^{-}   U_{\sigma} U^{-}   ~\}.\nn
\eeqa
By using Bianchi identity (\ref{newbianchi}) the last expression can
be reduced to the following boundary integral:
\beqa
\Pi(A^U) - \Pi(A)  &=& {i  \over   g}  \int_{ M_5}  d^5x
~\varepsilon^{\mu\nu\lambda\rho\sigma} ~Tr~ \{  \nabla_{\rho}
(G_{\mu\nu,\lambda}   U_{\sigma} U^{-}+
 G_{\mu\nu }  U_{\lambda} U^{-}  U_{\sigma} U^{-} )\}\nn\\
&=& {i  \over   g}  \int_{ M_5}  d^5x
~\varepsilon^{\mu\nu\lambda\rho\sigma} \partial_{\rho}  ~
Tr~  (G_{\mu\nu,\lambda}   U_{\sigma} U^{-}
+ G_{\mu\nu }  U_{\lambda} U^{-}  U_{\sigma} U^{-} ) \nn\\
&=& {i \over  g}  \int_{ \partial M_5}
~\varepsilon^{\mu\nu\lambda\rho\sigma}  ~Tr~  (G_{\mu\nu,\lambda}   U_{\sigma} U^{-}
+ G_{\mu\nu }  U_{\lambda} U^{-}  U_{\sigma} U^{-} )  ~d \sigma_{\rho}.\nn
\eeqa
Thus, the large gauge transformation of $\Pi(A)$ is
\beqa\label{largegaugetransformation1}
\Pi(A^U) - \Pi(A) =   {i  \over   g}  \int_{ M_5}  d^5x
~\varepsilon^{\mu\nu\lambda\rho\sigma} \partial_{\rho}  ~
Tr~  (G_{\mu\nu,\lambda}   U_{\sigma} U^{-}
+ G_{\mu\nu }  U_{\lambda} U^{-}  U_{\sigma} U^{-} ).
\eeqa
The expression (\ref{largegaugetransformation1}) is analogous to the corresponding
formula for  the Chern-Simons integral (\ref{largechernsimons}) and to our
previous result obtained  for the $\Sigma(A^U) - \Sigma(A)$ in (\ref{largegaugetransformation}).
We can introduce now a lower-dimensional functional representing potential anomaly in
four-dimensions:
\beqa\label{secondff}
\pi^1_4(A,U) =  ~\varepsilon^{\mu\nu\lambda\rho}
Tr~ (G_{\mu\nu,\lambda}   U_{\sigma} U^{-} ) ~.
\eeqa
The dimensional reduction of the variation $\delta \Sigma$
to $\sigma^1_{3}(A,U)$ of (\ref{secondf}) and of  $\delta \Pi$
to   $\pi^1_{4}(A,U)$ of (\ref{secondff})
stops  because a gauge transformation of the last densities leaves  their form unchanged.

\section{\it  Anomalies and  Densities in Higher Dimensions}
As we already discussed in the introduction, all chiral anomalies, Abelian
and non-Abelian, can be determined by a differential geometric method without having to evaluate a Feynman diagram
\cite{Zumino:1983ew,Stora:1983ct,Faddeev:1984jp,LBL-16443,Manes:1985df,Treiman:1986ep,
Faddeev:1987hg,AlvarezGaume:1985ex,Grimm:1974pr,Townsend:1983ana,AlvarezGaume:1983ig,Bardeen:1984pm,
AlvarezGaume:1983cs,AlvarezGaume:1984dr,AlvarezGaume:1985yb}.
The non-Abelian anomaly  in $ 2n$-dimensional space-time may be obtained  from
the Abelian anomaly in $2n+2$ dimensions by a reduction (transgression) procedure
\cite{Zumino:1983ew,Stora:1983ct,Faddeev:1984jp,LBL-16443,Manes:1985df,
Treiman:1986ep,Faddeev:1987hg,AlvarezGaume:1985ex}.
The $U_A(1)$ anomaly appears in
the divergence of the axial U(1) current $J^{A}_{\mu}= \bar{\psi} \gamma_{\mu}\gamma_5 \psi$
and in four-dimensional space-time it is given  by
\beqa\label{anomaly4d}
\partial^{\mu} J^{A}_{\mu} = -{1\over 16 \pi^2} \varepsilon^{\mu\nu\lambda\rho } Tr (G_{\mu\nu}G_{\lambda\rho})
=-{1\over 4 \pi^2}\varepsilon^{\mu\nu\lambda\rho } \partial_{\mu} ~Tr (A_{\nu}\partial_{\lambda}A_{\rho }
-i{2\over 3}g  A_{\nu}A_{\lambda}A_{\rho }).
\eeqa
The non-Abelian anomaly appears in the covariant divergence of the non-Abelian
left- and right-handed currents
$J^{aL}_{\mu}= \bar{\psi}_L \gamma_{\mu}\gamma_5 \sigma^a\psi_L$,
$J^{aR}_{\mu}= \bar{\psi}_R \gamma_{\mu}\gamma_5 \sigma^a\psi_R$ with
\beqa\label{nonabeliananomaly}
D^{\mu} J^{aL}_{\mu}
=-{1\over 24 \pi^2} \varepsilon^{\mu\nu\lambda\rho } \partial_{\mu} ~Tr [\sigma^a (A_{\nu}\partial_{\lambda}A_{\rho }
-i{1\over 2}g  A_{\nu}A_{\lambda}A_{\rho })].
\eeqa
The Abelian anomaly is gauge invariant, while the non-Abelian anomaly is gauge covariant and is
given by the covariant divergence of a non-Abelian current.

In order to introduce higher-dimensional densities it is convenient to use
the language of forms \cite{Zumino:1983ew,Stora:1983ct,Faddeev:1987hg}.
We already introduced the one- and two-form gauge potentials
$A=-ig A^{a}_{\mu} L_a dx^{\mu}$ and $A_2=-ig A^{a}_{\mu\nu} L_a dx^{\mu}dx^{\nu}$
with the corresponding field-strength tensors (\ref{fieldstrengthparticular}):
\be
G  = dA  + A^{2}  ~,~~~~G_3 = d A_2 +[A  , A_2].
\ee
The Bianchi identities (\ref{bianchi}) and (\ref{newbianchi}) now take the form
\be
DG  =0,~~~~DG_3 +[A_2, G ]=0,
\ee
where $DG  = dG  + [A , G ]$ and $DG_3 = dG_3 + A G_3 + G_3 A $.
Let us consider a higher-dimensional invariant density in $2n+3$ space-time dimensions:
\be
\Gamma_{2n+3} = Tr(G^n  G_3),
\ee
which coincides with $\Gamma(A)$  of (\ref{freeactionthreeprimesum}) for $n=1$ and is a natural generalization
of the Chern-Pontryagin form  $\CP_{2n}=Tr(G^n)$.
By direct computation of the derivative one can prove that $\Gamma_{2n+3}$ is an exact form:
\beqa
d \Gamma_{2n+3} &=& Tr(dG  G^{n-1}  G_3 +...+G^{n-1}  dG  G_3 + G^{n} dG_3)\nn\\
&=&Tr((dG  +[A ,G ]) G^{n-1}  G_3 +...  +
G^{n} (dG_3 +A G_3 +G_3 A ) ) \nn\\
&=& Tr(DG  G^{n-1}  G_3 +...+G^{n-1}  DG  G_3 + G^{n} DG_3)\nn\\
&=& Tr(G^{n} DG_3)= Tr(G^{n} (DG_3 +[A_2,G ]))=0. \nn
\eeqa
In this calculation one must change sign when transmitting the differential d through an odd form
or commuting odd forms using the cyclic property of the trace, and use Bianchi identities as well.
According to Poincar\'e's lemma, this equation implies that $\Gamma_{2n+3}$ can be
locally written as an exterior differential of a certain (2n +2)-form.
In order to find the form of which $\Gamma_{2n+3}$ is the derivative we have to find
its variation, induced by the variation of the fields $\delta A$ and
$\delta A_2$:
$$
\delta G = D (\delta A) ,~~~~\delta G_3 = D(\delta A_2) + [\delta A ,A_2]
$$
yielding a variation of $\Gamma_{2n+3}$ which is a total derivative:
\be\label{variation1}
\delta \Gamma_{2n+3} = d ~Tr(\delta A  G^{n-1}  G_3 +...+G^{n-1}  \delta A  G_3 + G^{n}  \delta A_2).
\ee

Following \cite{Zumino:1983ew}, we introduce
a one-parameter family of potentials and strengths  through the parameter t ($0\leq t \leq 1$):
\be
A_{t}= t A,~~~G_{t}= t G +(t^2-t)A^{2},~~~
A_{2t}= t A_2,~~~G_{3t}= t G_3 +(t^2-t)[A,A_{2}],
\ee
so that the equation (\ref{variation1}) can be rewritten as
$$
\delta Tr(G^n_t ~ G_{3t}) = d ~Tr(\delta A_t  G^{n-1}_t  G_{3t} +...+G^{n-1}_t  \delta A_t  G_{3t}
+ G^{n}_t  \delta A_{2t}).
$$
With $\delta = \delta t (\partial/ \partial t)$ and $\delta A_{t}=A \delta t$, $\delta A_{2t}=A_2 \delta t$
we shall get by integration the
desired result:
\be\label{result}
Tr(G^n  G_3) = d ~\sigma_{2n+2}~,
\ee
where the corresponding secondary $(2n+2)$-form is
\be\label{secondaryform}
\sigma_{2n+2}(A,A_2) =  \int^1_0 d t ~Tr(A  G^{n-1}_{t} G_{3t} +...+G^{n-1}_{t}  A G_{3t} + G^{n}_{t} ~ A_2).
\ee
The dimensionality of this density is $[mass]^{n(n+2)}$ and it can be used as an addition
to the (2n+2)-dimensional Lagrangian density  \cite{arXiv:1001.2808}
\be
{1 \over F^{n^2 -2}} \int_{M_{2n+2}} \sigma_{2n+2}(A,A_2),
\ee
where F is a dimensional coupling constant.
The $n=1$ case was considered in
\cite{arXiv:1001.2808}. It has the form:
$$
F \int_{M_{4}} \sigma_{4}(A,A_2)
$$
and can be added to the Lagrangian density generating masses of the vector gauge bosons.
The equation (\ref{secondaryform}) is a generalization of the formula
for the $n$-th Chern-Pontryagin character and
of the corresponding Chern-Simons secondary topological invariant $(2n-1)$-form
\cite{Zumino:1983ew,Stora:1983ct,Treiman:1986ep}
\footnote{The derivation can also be found in Appendix C.}:
\be
Tr(G^n) = d ~\omega_{2n-1},~~~~~
\ee
where
\be\label{zumino}
\omega_{2n-1}(A)= n \int^1_0 dt~  Tr (AG^{n-1}_t).
\ee
In $\CD = 2n$ dimensions, the $U_A(1)$ anomaly is given by this $2n$-form,  the
higher-dimensional analog of eq.~(\ref{anomaly4d}):
$$
d*J^A ~\propto ~Tr(G^n)= d ~\omega_{2n-1}.
$$
Note that the significance of the densities  (\ref{secondaryform}) in extended YM theory
and their connection with anomalies in different
dimensions is yet to be understood.

Returning back to the expression
(\ref{secondaryform}), for $n=1$ one can recover the  expressions
(\ref{freeactionthreeprimesum}) and
(\ref{topologicalcurrent1}) in four dimensions:
\be\label{sigma4}
\sigma_{4} =\int^1_0 d t ~Tr(A  G_{3t}  + G_{t} ~ A_2).
= Tr( G A_2)
\ee
In six dimensions, $n=2$, we have
$$
\sigma_{6} =\int^1_0 d t ~Tr(A  G_{t} G_{3t} +G_{t}  A G_{3t} + G^{2}_{t} ~ A_2),
$$
and after integration over t we get a secondary 6-form:
\beqa\label{newsecondary}
\sigma_{6}(A,G,G_3)&=&{1\over 3} Tr( A  G  G_3 + A  G_3 G   + A_2 G^{2}  - {1\over 2} A^3  G_3\nn\\
&-&{1\over 2} (A^2  A_2 -A A_2 A  + A_2 A^2 )G  + {1\over 2} A^4  A_2 ).
\eeqa
The new property of the last functional compared with $\sigma_4$ above is that when the
field-strength tensors tend to zero,
$G =G_3=0$,  the functional does not vanish and is equal to
\be\label{remnant}
 {1\over 6 } Tr(   A^4 A_2 ),
\ee
where one should substitute the flat connections (\ref{puregauge}) and (\ref{puregaugetensor}).
The subsequent forms $\sigma_{2n+2}$ are in $\CD=2n+2=4,6,8,10,\dots$ dimensions.
The form (\ref{newsecondary}) is the analog of the Chern-Simons (CS) 5-form~\cite{Treiman:1986ep}
\beqa
\omega_5(A,G)= Tr(A  G^2  - {1\over 2} A^3  G  + {1\over 10}A^5 ) ,\nn
\eeqa
and (\ref{remnant}) is the analog of the winding number $Tr(A^5 )$ in five dimensions
and of the corresponding  Wess-Zumino-Witten chiral effective action \cite{Witten:1983tw}.

The second series of invariant forms can be constructed by generalization of the
density  (\ref{invarinatinsixdimensions}) to higher dimensions.
It can be written as follows:
\be\label{result2}
\Delta_{6n}= Tr(G_3)^{2n},
\ee
and is an exact $6n$-form:
\beqa
d \Delta_{6n}&=& n ~Tr~G_3^{2n-2} (dG_3 G_3 - G_3 dG_3) = n~ Tr~ G_3^{2n-2}(DG_3 G_3 - G_3 DG_3)\nn\\
&=&n~ Tr~G_3^{2n-2} ((dG_3 +[A_2 G]) G_3 - G_3 (dG_3+ [A_2,G]))=0.
\eeqa
Its variation over the gauge fields is
\be
\delta \Delta_{6n} = ~2n~d ~ Tr (\delta A_2 G^{2n-1}_{3}) = d \pi_{6n-1},
\ee
so that after introducing the $t$ deformation of the fields   we get
the $(6n-1)$-form:
\be\label{secondaryg3}
\pi_{6n-1}(A,A_2) = 2n \int^{1}_{0} ~d t ~ Tr (A_2 G^{2n-1}_{3t}).
\ee
For $n=1$ it  reproduces the 5-form (\ref{topologicalSigmaCShigh})
\be
\pi_5 =2 \int^{1}_{0} ~d t ~ Tr (A_2 G_{3t})= Tr(A_2 G_3)
\ee
and for $n=2$ we get the 11-form
\be
\pi_{11} = 4 \int^{1}_{0} d t  ~Tr  A_2 ~(t G_3 +(t^2-t)[A,A_{2}])^3 ,
\ee
which after integration yields
\beqa
\pi_{11} =Tr   (~  A_2 (G^3_3 -{1\over 5}([A,A_{2}] G^{2}_{3} + G_{3}[A,A_{2}] G_{3}
+G^{2}_{3}[A,A_{2}]) + \nn \\
+{1\over 15}([A,A_{2}]^2 G_{3} +[A,A_{2}]  G_{3} [A,A_{2}]+ G_{3}[A,A_{2}]^2 )
-  {1\over 35} [A,A_{2}]^3 )~).
\eeqa
The forms $\pi_{6n-1}$ are defined in $\CD=6n-1=5,11,17,\dots$ dimensions.

Now, having in hand the $(2n+2)$-form $\sigma_{2n+2}$ (\ref{secondaryform})
and $(6n-1)$-form $\pi_{6n-1}$ (\ref{secondaryg3}),
we can turn to the construction of possible anomalies in extended YM theory. Indeed,
the non-Abelian anomalies can be determined through the geometric procedure as follows.
From the gauge invariance of the
densities $\Gamma_{2n+3}$ and $\Delta_{6n}$ and from (\ref{result}) and (\ref{result2}) we know that
\beqa
\delta ~\Gamma_{2n+3}=   \delta ~ d ~\sigma_{2n+2} = d ~\delta  ~ \sigma_{2n+2} =0,\nn\\
\delta ~\Delta_{6n}=  \delta  ~d ~\pi_{6n-1} = d ~\delta  ~ \pi_{6n-1} =0.\nn
\eeqa
Thus locally there must exist a certain $(2n+1)$-form  $\sigma^1_{2n+1}(\xi,A )$ and
$(6n-2)$-form $\pi^1_{6n-2}(\xi,A )$ such that
\be
\delta    \sigma_{2n+2}(A)= d\sigma^1_{2n+1}(\xi,A ) ,~~~~~~
\delta  \pi_{6n-1}(A) = d \pi^1_{6n-2}(\xi,A )~.
\ee
Here the  superscript of $\sigma^1_{2n+1}(\xi,A )$  and $\pi^1_{6n-2}(\xi,A )$
indicates that these forms are of first order in the gauge parameters.

These anomalies for lower dimensions are already known through the calculations we made in
the previous sections. Indeed, we calculated the global gauge variation of the
secondary forms $\sigma_4$ and $\pi_5$ and  found that they are total derivatives,
thus  from (\ref{secondf}) for infinitesimal
gauge transformations (\ref{small}) it follows that
\be
\sigma^1_3(\xi_1,A) = Tr(\xi_1 G),
\ee
where $\xi_1$ is a 1-form gauge parameter $\xi_1 = L^a \xi^a_{\mu} dx^{\mu}$,
and from (\ref{secondff})
we can extract $\pi^1_4$:
\be
\pi^1_4(\xi_1,A) = Tr(\xi_1 G_3).
\ee
In both cases there is no dependence from the scalar gauge parameter $\xi  = L^a \xi^a$.

In order to calculate the variation of the secondary characteristics in higher dimensions
we need the formulas for the gauge transformation of the various fields involved
in the expressions for $\sigma_{2n+2}$ (\ref{secondaryform})   and $\pi_{6n-1}$ (\ref{secondaryg3}),
which read
\beqa\label{gaugevariation}
&\delta_{\xi}A = d\xi + [A,\xi],~~~~~\delta_{\xi}A_2 = d\xi_1 + A\xi_1 +\xi_1 A + [A_2,\xi] ,~~~\nn\\
&\delta_{\xi}d A =  [d A,\xi]- A d\xi -d\xi A,~~~~~\delta_{\xi}d A_2 =  [d A,\xi_1]-
[A, d \xi_1]  + [d A_2, \xi] + [A_2, d \xi],~~~\nn\\
&\delta_{\xi} G_t = [G_t,\xi] +(t^2-t) (A d\xi  +d \xi  A),~\nn\\
&\delta_{\xi} G_{3t} = [G_{3t},\xi]+[G_t,\xi_1] +(t^2-t) ([A, d\xi_1]  +[A_2,d \xi]).
\eeqa
Let us calculate the variation of  $\sigma_6$ (\ref{newsecondary}) using the above formulas.
It turns out that there are many cancelations
between different terms, so that at the end we get two contributions, one linear in $d\xi$ and
the other - in $d\xi_1$. The term linear in the differential of the 1-form gauge parameter $d\xi_1$ is
\beqa
\delta \sigma_6 &=& Tr(d\xi_1 G^2  - {1\over 2} A^2 d\xi_1 G + {1\over 2} A  d\xi_1 A G
- {1\over 2} d\xi_1 A^2  G  +{1\over 2} A^4 d\xi_1) =\nn\\
&=&Tr(d\xi_1 ~d(A dA +{1\over 2} A^3))=d \sigma^1_5, \nn
\eeqa
therefore
\be\label{sigma511}
\sigma^1_{5}(\xi_1,A) =Tr(\xi_1 ~d(A dA +{1\over 2} A^3)).
\ee
This expression coincides with the standard gauge anomaly in four
dimensions $\omega^1_4$ (\ref{nonabeliananomaly}), with the only difference
that it is multiplied by $\xi_1$, which is here a 1-form.
The second term linear in $d\xi$ is
\beqa
\delta \sigma_6 &=& Tr(d\xi G G_3 +  d\xi  G_3 G  - {1\over 2}( d\xi A^2  G_3 + A d\xi A G_3 +
A^2 d\xi G_3) -\nn\\
&-& {1\over 2} (d\xi A A_2 G + A d\xi A_2 G - d\xi A_2 A G - A A_2 d\xi G + A_2 d\xi A G +A_2 A d\xi G)+\nn\\
&+&{1\over 2} (d\xi  A^3  A_2   + A d\xi  A^2  A_2   + A^2 d\xi  A   A_2   + A^3 d\xi   A_2  ) =\nn\\
&=&Tr\left(d\xi  ~d(A dA_2  +A_2 dA + {1\over 2}( A^2 A_2 - A A_2 A + A_2 A^2))\right)=d \sigma^1_5, \nn
\eeqa
and therefore
\beqa\label{sigma512}
\sigma^1_{5}(\xi,A,A_2) &=&Tr\left( \xi  ~d(A dA_2  +A_2 dA + {1\over 2}( A^2 A_2 - A A_2 A + A_2 A^2))\right)\nn\\
&=&Tr\left( \xi  ~d(A G_3  +A_2 G - {1\over 2}( A^2 A_2 - A A_2 A + A_2 A^2))\right).
\eeqa
The total form $\sigma^1_5$ is a sum of the two expressions above,  (\ref{sigma511}) and (\ref{sigma512}).
What is interesting here is that $\sigma^1_{5}$  explicitly contains the second-rank gauge field $A_2$
when we perform the standard YM infinitesimal gauge transformation $\xi$.
Because it is defined in odd
dimensions it may have contribution to the parity-violating anomaly \cite{AlvarezGaume:1984nf} and
its descendant ($\delta \sigma^1_5 = d \sigma^2_4$):
\be
\sigma^2_4(\xi,\eta,A) = Tr\left( (d \xi~ \eta + \eta~ d \xi -\xi ~d \eta - d\eta~ \xi )~ d A_2 \right)
\ee
may represent a potential anomalous Schwinger term  in the corresponding
gauge algebras  \cite{Faddeev:1984jp}. Here the  superscript of $\sigma^2_4(\xi,\eta,A)$
indicates that this form is of second order in the gauge parameters.

\section{\it Conclusions}
In conclusion  let us compare the Pontryagin-Chern-Simons densities
$\CP_{2n}$, $\omega_{2n-1}$ and $\omega^{1}_{2n-2}$ in
YM gauge theory with the corresponding  densities
$\Gamma_{2n+3}$, $\sigma_{2n+2}$, $\sigma^{1}_{2n+1}$
and $\Delta_{6n}$, $\pi_{6n-1}$, $\pi^{1}_{6n-2}$  in the extended YM theory.
The new characteristic classes are local forms defined on the space-time manifold
and constructed from the curvature 2-form $G$ and 3-form $G_3$:
\beqa
\Gamma_{2n+3} = Tr(G^n  G_3) = d ~\sigma_{2n+2}~,~~~~~~~~
\Delta_{6n}=  Tr(G_3)^{2n}= d~\pi_{6n-1}~.
\eeqa
These characteristic classes are closed forms, but not globally exact.
The secondary characteristic classes $\sigma_{2n+2}$ and $\pi_{6n-1}$
can be expressed in integral form (\ref{secondaryform}) and (\ref{secondaryg3}) in analogy with the
Chern-Simons form (\ref{zumino}). Their gauge variation can also be found,
yielding the potential anomalies in gauge field theory. The above general considerations
should be supplemented  by an explicit calculation of loop diagrams involving chiral fermions.
The argument in favor of the existence of
these potential anomalies is based on the fact that they fulfill Wess-Zumino consistency conditions.
At the same time, these invariant densities constructed on the space-time manifold
have their own independent value since they suggest the existence of  new
invariants characterizing topological properties of a manifold.

The Abelian version of the invariant $\Sigma(A)$ was investigated earlier
in \cite{Cremmer:1973mg,Kalb:1974yc,Nambu:1975ba,Aurilia:1981xg,
Freedman:1980us,Rohm:1985jv,Allen:1990gb,Slavnov:1988sj,Giddings:1987cg,Bowick:1988xh}.
Attempts to construct a non-Abelian  invariant
in a similar way  have come up with difficulties  because they
involve non-Abelian generalization of
gauge transformations of antisymmetric fields
\cite{Henneaux:1997mf,Thierry-Mieg,Cantcheff:2011qu}.
The no-go theorem \cite{Henneaux:1997mf} implies that without additional
auxiliary fields the gauge transformations cannot form a closed group.
And, indeed, the gauge transformations of non-Abelian tensor gauge fields
$\delta_{\xi}$   (\ref{polygauge})  cannot be limited to a YM 1-form and rank-2 antisymmetric
field. Instead, the antisymmetric tensor is
extended by a symmetric rank-2 gauge field, so that together they form a
gauge field $A^{a}_{\mu\nu}$ which transforms according  to
(\ref{polygauge}) and is a fully propagating field.
It is also important that one should include all high-rank gauge fields $
A^{a}_{\mu\lambda_1 ... \lambda_{s}} $ in order to be able
to close the group of gauge transformations and to construct an
invariant Lagrangian.

\section*{\it Acknowledgements}
One of us G.S. would like to thank Ludwig Faddeev, Luis Alvarez-Gaume and Raymond Stora
for stimulating discussions and   CERN Theory Division, where part of this work was completed,
for hospitality. This work was supported in part by the European Commission
under the ERC Advanced Grant 226371 and the contract PITN-GA-2009-237920.

%\vskip 0.5cm
\appendix %{\Large\it Appendix A. Tensor gauge fields}
%\vskip 0.2cm

\section{\it Tensor gauge fields}

The extended non-Abelian gauge transformation $
\delta_{\xi} $ of  tensor gauge fields
is defined
by the equations \cite{Savvidy:2005fi,Savvidy:2005zm,Savvidy:2005ki}:
\beqa\label{polygauge}
\delta_{\xi}  A_{\mu} &=& \partial_{\mu}\xi -i g[A_{\mu},\xi]\nonumber\\
\delta_{\xi}  A_{\mu\nu} &=& \partial_{\mu}\xi_{\nu} -i g[A_{\mu},\xi_{\nu}]
-i g [A_{\mu\nu},\xi]\nonumber\\
\delta_{\xi}  A_{\mu\nu\lambda} &=& \partial_{\mu}\xi_{\nu\lambda}
-i g[A_{\mu},\xi_{\nu\lambda}]-
i g[A_{\mu\nu},\xi_{\lambda}]-i g [A_{\mu\lambda},\xi_{\nu}]
-i g [A_{\mu\nu\lambda},\xi],\\[-7pt]
 &\cdot&\nn\\[-17pt]
 &\cdot&\nn\\[-17pt]
 &\cdot&\nn
%&~&..............................,\nn
\eeqa
where $\xi^{a}_{\lambda_1 ... \lambda_{s}}(x)$ are totally symmetric gauge parameters,
and comprises a closed algebraic structure.
The tensor gauge fields are in the matrix representation
$A^{ab}_{\mu\lambda_1 ... \lambda_{s}} =
(L_c)^{ab}  A^{c}_{\mu\lambda_1 ... \lambda_{s}} = i f^{acb}A^{c}_{\mu
\lambda_1 ... \lambda_{s}}$  with
$f^{abc}$ - the structure constants. The generalized field-strength tensors
(\ref{fieldstrengthparticular}) transform homogeneously
under the extended gauge transformations $\delta_{\xi} $:
\beqa\label{fieldstrengthtensortransfor}
\delta G^{a}_{\mu\nu}&=&  -i g  [G_{\mu\nu} ~\xi] , \\
\delta G^{a}_{\mu\nu,\lambda} &=& -i g  (~[~G_{\mu\nu,\lambda}~ \xi ]
+  [G_{\mu\nu} ~\xi_{\lambda}]~),\nonumber\\
\delta G^{a}_{\mu\nu,\lambda\rho} &=& - i g
(~[G^{b}_{\mu\nu,\lambda\rho} ~\xi]
+ [ G_{\mu\nu,\lambda} ~\xi_{\rho}] +
[G_{\mu\nu,\rho} ~\xi_{\lambda}] +
[G_{\mu\nu} ~\xi_{\lambda\rho}]~),\nn\\[-7pt]
 &\cdot&\nn\\[-17pt]
 &\cdot&\nn\\[-17pt]
 &\cdot&\nn
% ......&.&............................................\nn
\eeqa
In the YM theory the Bianchi identity is
\be\label{bianchi}
[\nabla_{\mu},G_{\nu\lambda}]+[\nabla_{\nu},G_{\lambda\mu}]+
[\nabla_{\lambda},G_{\mu\nu}]=0,
\ee
and for the higher-rank field-strength tensors $G_{\nu\lambda,\rho}$ and
$G_{\nu\lambda,\rho\sigma}$  the Bianchi identities are:
\be\label{newbianchi}
[\nabla_{\mu},G_{\nu\lambda,\rho}]-ig[A_{\mu\rho},G_{\nu\lambda}]+
[\nabla_{\nu},G_{\lambda\mu,\rho}]-ig[A_{\nu\rho},G_{\lambda\mu}]+
[\nabla_{\lambda},G_{\mu\nu,\rho}]-ig[A_{\lambda\rho},G_{\mu\nu}]=0,
\ee
\be\label{newbianchi3}
[\nabla_{\mu},G_{\nu\lambda,\rho\sigma}]-ig[A_{\mu\rho},G_{\nu\lambda,\sigma}]
-ig[A_{\mu\sigma},G_{\nu\lambda,\rho}]-ig[A_{\mu\rho\sigma},G_{\nu\lambda}]
+ cyc.perm. (\mu\nu\lambda)=0
\ee
and so on.

\vskip 0.2cm

\section{\it Winding Number}
The variation of the CS term under large gauge transformation can be computed by
using  parametrization (\ref{parametrization}) for the SU(2) group elements
\cite{Jackiw:1979ur,Jackiw:2004fg,Jackiw:2005wp}. We used the mathematica program
"Exterior Differential Calculus" developed
by Sotirios Bonanos \cite{sotirios} to evaluate the wedge products
\beqa
W(A^U) - W(A)&=&
{1\over 24 \pi^2} \int_{M_3}  d^3x ~\varepsilon^{ijk }~
Tr~ (U^{-}\partial_{i}U ~U^{-}\partial_{j}U ~U^{-}\partial_{k}U  )\nn\\
&=&
{1\over 24 \pi^2} \int_{M_3}   ~
Tr~ (U^{-}d U \wedge U^{-}d U \wedge U^{-} d U  )=\nn\\
&=&{1  \over 2  \pi^2} \int_{M_3}  ~{\sin^2\vert \xi \vert \over \vert \xi \vert^2}~
 d \xi_1 \wedge d\xi_2 \wedge d\xi_3  =\nn\\
&=&-{1  \over 16 \pi^2}  \int_{M_3}  \varepsilon^{abc}
d~(\hat{\xi}^a  ~ \hat{\xi}^b \wedge d\hat{\xi}^c ~
(\sin 2\vert \xi \vert -2 \vert \xi \vert) )=\nn\\
&=&-{1  \over 16 \pi^2} \int_{\partial M_3}  \varepsilon^{abc}
 ~ \hat{\xi}^a  ~ \hat{\xi}^b \wedge d\hat{\xi}^c ~
(\sin 2\vert \xi \vert -2 \vert \xi \vert)=\nn\\
&=&-{1  \over 8 \pi^2}  \int_{\partial M_3} (\sin 2\vert \xi \vert -2 \vert \xi \vert)
 ~ \hat{\xi}^a    d\omega^a ~
,
\eeqa
where $d\omega^a = 2 \varepsilon^{abc}  ~ d \hat{\xi}^b \wedge d\hat{\xi}^c$.~
With the boundary condition
$
\vert \xi \vert ~~~{\over r \rightarrow \infty}> ~~~\pi N
$
one gets
\beqa
W(A^U) - W(A)&=&{1  \over 4 \pi^2} \int_{\partial M_3}
 ~\vert \xi \vert  \hat{\xi}^a   d\omega^a ~
 = N.
\eeqa

%\vskip 0.2cm
%\appendix{\Large\it Appendix C. Non-Abelian Anomaly}
%\vskip 0.2cm

\section{\it Appendix C. Non-Abelian Anomaly}

As an exercise let us
calculate the non-Abelian anomaly by gauge variation of $\omega_{2n-1}$ in YM theory
(\ref{zumino}) \cite{Zumino:1983ew,Stora:1983ct,Faddeev:1984jp,LBL-16443,Manes:1985df,Treiman:1986ep,Faddeev:1987hg}. Using formulas (\ref{gaugevariation}) one can get
\beqa
\delta_{\xi}\omega_{2n-1}(A,G)= n \int^{1}_{0} dt ~\delta_\xi ~Str (A G^{n-1}_{t})
=~~~~~~~~~~~~~~~~~~~~~~~~~~~~~~~~~~~~~~~~~~~\nn\\
=n \int^{1}_{0} d t ~Str ( ~(d\xi +[A,\xi]) G^{n-1}_{t}
+ A ([G_t,\xi] +(t^2-t)\{A,d\xi\})G^{n-2}_{t} +\nn\\
+A G_{t} ([G_t,\xi] +(t^2-t)\{A,d\xi\})G^{n-3}_{t} +...+AG^{n-2}_{t}([G_t,\xi] +(t^2-t)\{A,d\xi\}) ~ ). \nn
\eeqa
Collecting terms which contain $d\xi$ and $\xi$ into two separate integrals we get
\beqa
n \int^{1}_{0} d t ~Str( d\xi   G^{n-1}_{t} + (t^2-t)  A  (\{A,d\xi\} G^{n-2}_{t}
+ G_{t}  \{A,d\xi\} G^{n-3}_{t} +...+ G^{n-2}_{t} \{A,d\xi\}) ) \nn\\
+n \int^{1}_{0} d t ~Str( [A,\xi]   G^{n-1}_{t} +    A [G_t, \xi] G^{n-2}_{t}
+A G_{t}  [G_t, \xi] G^{n-3}_{t} +...+AG^{n-2}_{t} [G_t, \xi] ). \nn
\eeqa
Opening the brackets in the second integral one can see that it vanishes, while the first integral
can be represented as
\beqa
 n \int^{1}_{0} d t ~Str(  d \xi   G^{n-1}_{t} + (t^2-t) (n-1) ( \{A,A\} d\xi G^{n-2}_{t}
+   A  d \xi  [A,  G^{n-2}_{t}] )  ,\nn
\eeqa
and by using the equations
$$
dG^{n-2}_{t}= -[A_t,G^{n-2}_t],~~~~{\partial G_t \over \partial t}=dA +t \{A,A\}
$$
it can be rewritten as
\beqa
 n \int^{1}_{0} d t ~Str\left( d \xi   G^{n-1}_{t} + (t -1) (n-1) ( ({\partial G_t \over \partial t}- dA  )
     d\xi G^{n-2}_{t} -  A  d \xi  dG^{n-2}_{t} )\right). \nn
\eeqa
Integration by parts cancels the first and the second terms in the integrand, so that
\beqa
 n(n-1) \int^{1}_{0} d t  (t -1)  Str\left( -  dA
     d\xi G^{n-2}_{t} -  A  d \xi  dG^{n-2}_{t} )\right) =~~~~~~~~~~~~~~~~~~~~~\nn\\
=n(n-1) \int^{1}_{0} d t  (1-t) d  ~Str\left(\xi d ( G^{n-2}_{t}A) \right),      \nn
\eeqa
and we arrived to the celebrated result for the non-Abelian anomaly
\cite{Zumino:1983ew,Stora:1983ct,Faddeev:1984jp,LBL-16443,Manes:1985df,Treiman:1986ep,Faddeev:1987hg}:
\beqa
\omega^{1}_{2n-2}=n(n-1) \int^{1}_{0} d t  (1-t)  ~Str\left(\xi d (A G^{n-2}_{t}) \right)  .
\eeqa
In $\CD = 2n-2$ dimensions   non-Abelian anomaly is given by this $(2n-2)$-form,  the higher
dimensional analog of the eq.~(\ref{nonabeliananomaly}):
\be
D*J^A_{\xi} ~\propto    ~\omega^{1}_{2n-2}(\xi,A).
\ee

\vfill
\end{document}